\documentclass[conference]{IEEEtran}
\usepackage[linesnumbered,lined]{algorithm2e}
\usepackage{multirow}
\usepackage{booktabs}
\usepackage{caption}
\usepackage[english]{babel}
\usepackage[dvipsnames]{xcolor}
\usepackage{soul}
\usepackage{hyperref}

\IEEEoverridecommandlockouts
\usepackage{cite}
\usepackage{amsmath,amssymb,amsfonts}
\usepackage{algorithmic}
\usepackage{graphicx}
\usepackage{textcomp}
\usepackage{xcolor}
\def\BibTeX{{\rm B\kern-.05em{\sc i\kern-.025em b}\kern-.08em
    T\kern-.1667em\lower.7ex\hbox{E}\kern-.125emX}}
\usepackage{lipsum}

\newtheorem{thm}{Theorem}[section]

\newtheorem{proof}{Proof}

\newtheorem{defn}{Definition}[section]

\begin{document}

\newcommand{\swallow}[1]{}
\newcommand{\lastcuts}[1]{}
\newcommand{\todo}[1]{{\bf To do:} #1}
\newcommand{\SkienaA}[0]{$S_A$}
\newcommand{\figurethathastogo}[1]{}
\newcommand{\added}[1]{\textcolor{black}{#1}}
\newcommand{\removed}[1]{\textcolor{red}{\st{#1}}}
\newcommand{\blfootnote}[1]{
    \begingroup
    \renewcommand\thefootnote{}\footnote{#1}
    \addtocounter{footnote}{-1}
    \endgroup
}
\newcommand{\putlink}[0]{\url{https://github.com/aamgalan/spatial_autocorrelation}}

\title{Fast Spatial Autocorrelation}

\author{
\IEEEauthorblockN{Anar Amgalan}
\IEEEauthorblockA{\textit{Dept. of Physics and Astronomy} \\
\textit{Stony Brook University}\\
anar.amgalan@stonybrook.edu}
\and
\IEEEauthorblockN{LR Mujica-Parodi}
\IEEEauthorblockA{\textit{Dept. of Biomedical Engineering} \\
\textit{Stony Brook University}\\
lilianne.strey@stonybrook.edu}
\and
\IEEEauthorblockN{Steven S. Skiena}
\IEEEauthorblockA{\textit{Dept. of Computer Science} \\
\textit{Stony Brook University}\\
skiena@cs.stonybrook.edu}
}

\maketitle

\begin{abstract}

Physical or geographic location proves to be an important feature in many data science models, because many diverse natural and social phenomenon have a spatial component.
{\em Spatial autocorrelation} measures the extent to which locally adjacent observations of the same phenomenon are correlated.
Although statistics like Moran's $I$ and Geary's $C$ are widely used to measure spatial autocorrelation, they are slow: all popular methods run in $\Omega(n^2)$ time, rendering them unusable for large data sets, or long time-courses with moderate numbers of points.

We propose a new $S_A$ statistic based on the notion that the variance observed when merging pairs of nearby clusters should increase slowly for spatially autocorrelated variables.
We give a linear-time algorithm to calculate $S_A$ for a variable with an input agglomeration order (available at \putlink{}).
For a typical dataset of $n \approx 63,000$ points, our $S_A$ autocorrelation measure can be computed in 1 second, versus 2 hours or more for Moran's $I$ and Geary's $C$.
Through simulation studies, we demonstrate that $S_A$ identifies spatial correlations in variables generated with spatially-dependent model half an order of magnitude earlier than either Moran's $I$ or Geary's $C$.
Finally, we prove several theoretical properties of $S_A$: namely that it behaves as a true correlation statistic, and is invariant under addition or multiplication by a constant.

\end{abstract}


\section{Introduction}

\blfootnote{
    \added{
        ACKNOWLEDGMENTS. The research described in this paper was partially funded by the NSF (IIS-1926751, IIS-1927227, and IIS-1546113 to S.S.S.), the W. M. Keck Foundation (L.R.M.-P.) and the White House Brain Research Through Advancing Innovative Technologies (BRAIN) Initiative (NSFNCS-FR 1926781 to L.R.M.-P.).
    }
}

Physical or geographic location proves to be an important feature in many data science models, because many diverse natural and social phenomenon have a spatial component.
Geographic features such as longitude/latitude, zip codes, and area codes are often used in predictive models to capture spatial associations underlying properties of interest.
Some of this is for physical reasons: the current temperature at location $p_1$ is likely to be similar to that at $p_2$ if $p_1$ is near $p_2$, and the synchrony between two regions in the brain is a function of the network of physical connections between them.
But social and economic preferences in what people like, buy, and do also have a strong spatial component, due to cultural self-organization (homophily) as well as differential access to opportunities and resources.

Correlation measures (including the Pearson and Spearman correlation coefficients) are widely used to measure the degree of association between pairs of variables $X$ and $Y$.
By convention, the $corr(X,Y) = 0$ signifies that $X$ and $Y$ are independent of each other\lastcuts{, values $0 < corr(X,Y) \leq  1$ denote positive dependence on each other and $-1 \leq corr(X,Y) < 0)$ signify inverse dependencies}.
The strength of dependency, and our ability to predict $X$ given $Y$, increases with $|corr(X,Y)|$.
{\em Autocorrelation} of time series or sequential data measures the degree of association of $z_i$ and sequence elements with a lag-$l$, i.e.\ $z_{i+l}$.
{\em Spatial autocorrelation} measures the extent to which locally adjacent observations of the same phenomenon are correlated.

Spatial autocorrelation proves more complex to measure than sequence autocorrelation, because the association is multi-dimensional and bi-directional.
Social scientists and geoscience researchers have developed a rich array of statistics which endeavor to measure the spatial correlation of a variable $Z$, including Moran's $I$ \cite{moran1950}, Geary's $C$ \cite{geary1954contiguity}, and Matheron variogram \cite{matheron1963principles}.
For example, political preferences are generally spatially autocorrelated, as reflected by the notion of ``Red'' states and ``Blue'' states in the U.S.
There is a general sense that political preferences are increasingly spatially concentrated.
Spatial autocorrelation statistics provide the right tool to measure the degree to which this and related phenomena may be happening.

These statistics are widely used, particularly Moran's $I$ and Geary`s $C$, yet our experience with them has proven disappointing.
First, they are slow: all popular methods run in $\Omega(n^2)$ time, rendering them unusable for large data sets, or long time-courses with moderate numbers of points.
Second, although they are effective at distinguishing spatial correlated variables from uncorrelated variables from relatively few samples, they appear less satisfying in comparing the degree of spatial association among sets of variables.
\added{Other inroads to efficient spatial data analysis primarily concern with detection of outliers and anomalies \cite{1250986,1410286}.} 
In this paper, continuing the naming tradition of Moran's I and Geary's C, we humbly propose a new spatial autocorrelation statistic: Skiena's A or \SkienaA. 
We will primarily consider a dataset of 47 demographic and geospatial variables, measured over roughly 3,000 counties in the United States \cite{centers2017brfss,bureau2010profile,centers2018multiple,grieco2012foreign,mantel1967detection}, with results reported in Table \ref{tab:median-sorted}.
\added{The dataset was previously used in identification of socio-demographic variables determining county level substance abuse statistics in the U.S. \cite{curtis2018can}. }
With our preferred statistic, the median-clustered \SkienaA, the six geophysical variables measuring sunlight, temperature, precipitation, and elevation all scored as spatially autocorrelated above $0.928$, whereas the strongest demographic correlation (other language) came in at $0.777$, reflecting the concentration of Hispanic-Americans in the Southwestern United States.
\swallow{
    In contrast, Geary's $C$ found no less than eight demographic variables more spatially correlated than at least one geospatial variable, while Moran's $I$ found the log population density more spatially correlated than maximum temperature or the degree of sunlight.
}

\begin{figure*}[htbp]
  \includegraphics[width=\textwidth]{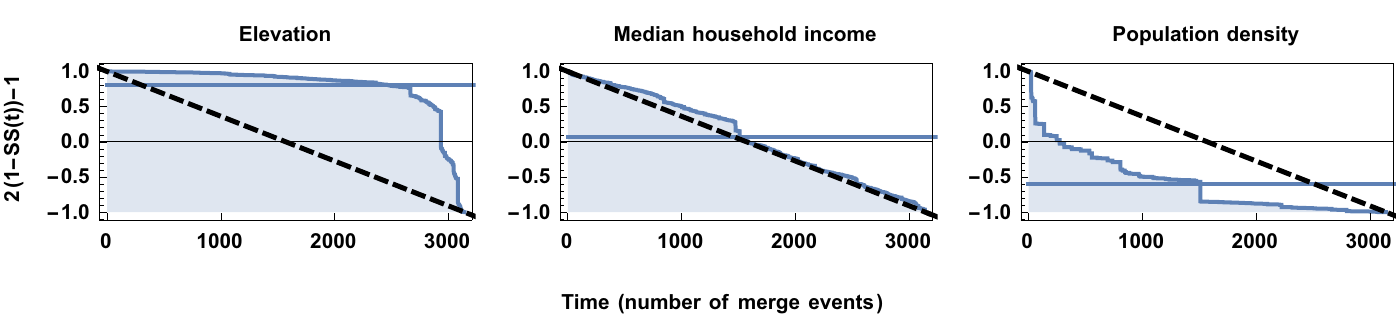}
  \caption{Representative traces of the single-linkage \SkienaA{} statistic (sum of squared deviations  $SS(t)$ scaled with $L(x)=2(1-x)-1$ to be in range [-1, 1]) as a function of the number of merging events, for selected U.S. county variables. 
  The area under the curve shows {\em Elevation} as strongly spatially correlated ({\bf \SkienaA{}=0.802}), {\em Median Income} as uncorrelated ({\bf \SkienaA{}=0.073}), and {\em Population Density} as spatially anti-correlated ({\bf \SkienaA{}=-0.598}).
  }
  \label{fig:trace}
\end{figure*}

\swallow{
    \begin{figure*}[htbp]
        \begin{center}
            \includegraphics[width=0.9\textwidth]{plot-pub-75_spatial_county_coordinates_3D_Mollweide_3-variables.pdf}
        \end{center}
        \caption{\added{The spatial distributions of the demographic variables (U.S. counties dataset). }
        }
        \label{fig:mollweide}
    \end{figure*}
}

Our statistic is based on the notion that spatially autocorrelated variables should exhibit low variance within natural clusters of points.
In particular, we expect the variance observed when merging pairs of nearby clusters should increase less the more spatially autocorrelated the variable is.
The within-cluster sum of squares of single points is zero, while the sum of squares of the single cluster after complete agglomerative clustering is $(n-1)\sigma^2$.
The shape of this trajectory from $0$ to $(n-1)\sigma^2$ after $n-1$ merging operations defines the degree of spatial autocorrelation, as shown in Fig.~\ref{fig:trace}.

Our major contributions in this paper include:

\begin{itemize}

\item
{\em Linear-time spatial correlation} --
The complexity to calculate \SkienaA{} for a variable defined by $n$ points and an input agglomeration order is $O(n)$, where traditional measures such as Moran's $I$ and Geary's $C$ require quadratic time.
This matters: for a typical dataset of $n \approx 63,000$ points, our \SkienaA{} autocorrelation measure can be computed in 1 second, vs. 2 hours for Moran's $I$ and Geary's $C$. 
Times shown are in seconds. 

\vspace{2mm}

\begin{tabular}{ l l l l l l }
    \hline
     & \multicolumn{4}{c}{Number of data points} \\
    statistic & 100 & 1000 & 10000 & 39810 & 63095 \\
    \hline
    Moran I & $\leq 1$ & $\leq 1$ & 60 & 1036 & 6784 \\
    Geary C & $\leq 1$ & 2 & 169 & 3112 & 11901 \\
    \SkienaA{} single & $\leq 1$ & $\leq 1$ & $\leq 1$ & $\leq 1$ & $\leq 1$ \\
    \SkienaA{} median & $\leq 1$ & $\leq 1$ & $\leq 1$ & $\leq 1$ & $\leq 1$ \\
    \hline
\end{tabular}

\vspace{2mm}

For points in two dimensions, the single-linkage agglomeration order can be computed in $O(n \log n)$.
Constructing more robust agglomeration orders like median-linkage may take quadratic time, however this computation needs to be performed only once when performing spatial analysis over $m$ distinct variables or time points.

We demonstrate the practical advantages of this win in an application on a brain fMRI time series data -- analyzing the results of a dataset roughly 36,000 times faster than possible with either Moran's I or Geary's C, had they not run out of memory in the process.
%

\item
{\em Greater sensitivity than previous methods} --
We assert that the median-clustered \SkienaA{} captures spatial correlations at least as accurately as previous statistics.
Through simulation studies, we demonstrate that it identifies spatial correlations in variables generated with spatially-dependent model half an order of magnitude earlier than either Moran's $I$ or Geary's $C$ (Fig.~\ref{fig:disk-mean}).
On the U.S. county data, we show that median-clustered \SkienaA{} correlates more strongly with Geary's $C$ (-0.943) and comparably with Moran's $I$ (0.879) than they do with themselves (-0.922).

\item
{\em Theoretical analysis of statistical properties} --
We demonstrate a variety of theoretical properties concerning \SkienaA.
We prove that it behaves as a true correlation statistic, ranging from $[-1,1)$ with an expected value of 0 for any $\textit{i.i.d.}$ random variable generated independent of location.
We show that $S_A(X) = S_A(a+X) = S_A(a \cdot X)$, meaning it is invariant under addition or multiplication by a constant.
Further, we show that \SkienaA{} measures increased spatial correlation as the sampling density increases, as should be the case for samples drawn from smooth functions -- but is not true for either Moran's $I$ or Geary's $C$.

\end{itemize}

The implementation of our statistic is available at \putlink{}.
This paper is organized as follows.
Section \ref{previous-work} introduces previous work on spatial autocorrelation statistics, with descriptions of six such statistics including the popular Moran's $I$ and Geary's $C$.
Our new \SkienaA{} agglomerative clustering statistic, with a fast algorithm to compute it, is presented in Section \ref{sec:method}.
Theoretical and experimental results are presented in Sections \ref{sec:analysis} and \ref{sec:evaluation}, respectively.

\begin{table*}
\centering
\caption{
Spatial autocorrelation for 47 geophysical and demographic variables on U.S. counties, sorted by their median-clustered \SkienaA{} value.
We note that the median-linkage agglomeration order produced the most satisfying ranking of variables by spatial autocorrelation compared to classical statistics and the weaker single-linkage aggregation order.
Median-clustered \SkienaA{} ranks all geophysical variables as more spatially autocorrelated than any demographic variable, and exhibits a stronger correlation with Geary's $C$ (-0.943) and comparable with Moran's $I$ (0.879) than they do with themselves (-0.922). 
For both \SkienaA{} metrics, the agglomeration order was computed only once and reused for all variables.
}
\label{tab:median-sorted}
    \begin{tabular}{ r c c c c c c }
        \cmidrule(r){1-6}
         & & Geary & Moran & \multicolumn{2}{c}{\SkienaA} & \multirow{40}{*}{
         \begin{minipage}{0.25\textwidth}
            \includegraphics[width=1.75in]{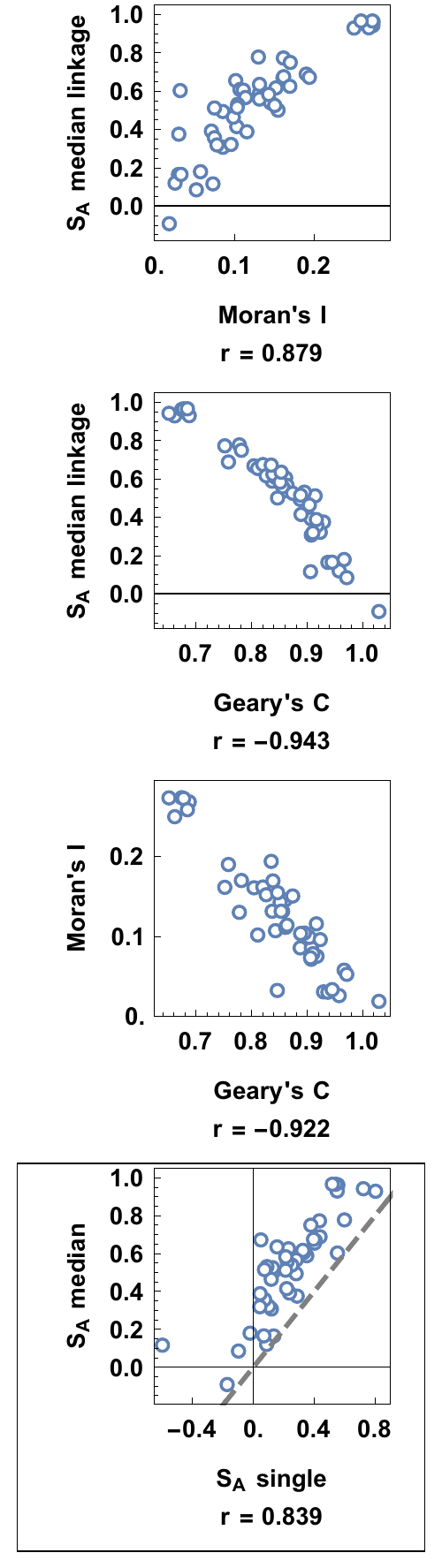}\captionof{figure}{Scatter of variables Moran's $I$, Geary's $C$, and $S_A$. U.S. counties dataset.}
            \label{fig:scatter}
        \end{minipage}
        } \\ 
        variable & n & C & I & single & median & \\ 
        \cmidrule(r){1-6}
        maxtemp & 3106 & 0.678 & 0.272 & 0.540 & 0.966 & \\
        sunlight & 3106 & 0.684 & 0.258 & 0.519 & 0.965 & \\
        mintemp & 3106 & 0.674 & 0.273 & 0.555 & 0.962 & \\
        precipitation & 3106 & 0.651 & 0.273 & 0.722 & 0.942 & \\
        max heat index & 3106 & 0.688 & 0.268 & 0.550 & 0.930 & \\
        elev & 3142 & 0.662 & 0.250 & 0.802 & 0.928 & \\
        \cmidrule(r){1-6}
        other language & 3142 & 0.778 & 0.130 & 0.598 & 0.777 & \\
        med house val & 3141 & 0.752 & 0.161 & 0.434 & 0.772 & \\
        log med house val & 3141 & 0.782 & 0.170 & 0.378 & 0.749 & \\
        log pop density & 3141 & 0.758 & 0.190 & 0.438 & 0.688 & \\
        main protestant & 3113 & 0.820 & 0.162 & 0.395 & 0.675 & \\
        percent black & 3142 & 0.835 & 0.194 & 0.049 & 0.672 & \\
        rep sen 2010 & 2115 & 0.804 & 0.161 & 0.406 & 0.668 & \\
        foreign born & 3142 & 0.811 & 0.102 & 0.401 & 0.654 & \\
        percent white & 3142 & 0.853 & 0.131 & 0.155 & 0.635 & \\
        evan protestant & 3122 & 0.838 & 0.169 & 0.231 & 0.625 & \\
        percent physically inactive chr & 3137 & 0.826 & 0.152 & 0.325 & 0.616 & \\
        rep pre 2012 & 3128 & 0.843 & 0.107 & 0.332 & 0.608 & \\
        catholic & 2958 & 0.861 & 0.111 & 0.330 & 0.605 & \\
        total pop & 3142 & 0.846 & 0.032 & 0.552 & 0.602 & \\
        percent obese chr & 3137 & 0.837 & 0.131 & 0.350 & 0.588 & \\
        high school & 3142 & 0.851 & 0.143 & 0.213 & 0.582 & \\
        rep pre 2008 & 3112 & 0.864 & 0.114 & 0.281 & 0.566 & \\
        year potential life lost rate chr & 2861 & 0.856 & 0.131 & 0.222 & 0.558 & \\
        percent excessive drinking chr & 2591 & 0.860 & 0.145 & 0.252 & 0.539 & \\
        log med house income & 3141 & 0.895 & 0.104 & 0.089 & 0.531 & \\
        percent fair or poor chr & 2738 & 0.874 & 0.150 & 0.126 & 0.525 & \\
        med house income & 3141 & 0.888 & 0.104 & 0.073 & 0.515 & \\
        rep hou 2010 & 3091 & 0.914 & 0.075 & 0.212 & 0.512 & \\
        separated & 3142 & 0.847 & 0.154 & 0.130 & 0.500 & \\
        motorvehicle mortality rate chr & 2828 & 0.888 & 0.086 & 0.281 & 0.493 & \\
        below poverty & 3141 & 0.904 & 0.099 & 0.117 & 0.464 & \\
        percent smokers chr & 2502 & 0.889 & 0.103 & 0.219 & 0.414 & \\
        divorced & 3142 & 0.907 & 0.071 & 0.237 & 0.391 & \\
        physically unhealthy days chr & 2954 & 0.917 & 0.116 & 0.044 & 0.388 & \\
        med age & 3142 & 0.930 & 0.031 & 0.287 & 0.375 & \\
        bac & 3142 & 0.918 & 0.075 & 0.074 & 0.357 & \\
        mentally unhealthy days chr & 2953 & 0.924 & 0.096 & 0.100 & 0.321 & \\
        grad & 3142 & 0.910 & 0.078 & 0.043 & 0.319 & \\
        married & 3142 & 0.908 & 0.086 & 0.118 & 0.307 & \\
        agasltrate & 2056 & 0.967 & 0.058 & -0.021 & 0.179 & \\
        ls 10 avg & 2004 & 0.945 & 0.033 & 0.069 & 0.165 & \\
        percent male & 3142 & 0.937 & 0.030 & 0.134 & 0.164 & \\
        same sex & 3142 & 0.958 & 0.026 & 0.087 & 0.120 & \\
        pop density & 3141 & 0.906 & 0.073 & -0.598 & 0.116 & \\
        robberyrate & 2056 & 0.971 & 0.052 & -0.098 & 0.085 & \\
        murderrate & 2056 & 1.029 & 0.019 & -0.172 & -0.092 & \\
        \cmidrule(r){1-6}
    \end{tabular}
\end{table*}

\section{Previous Work}\label{previous-work}

\subsection{Moran's $I$}

The most well-known of spatial autocorrelation metrics, Moran's $I$ \cite{moran1950} has been around for more than 50 years. 
Originally proposed as a way of capturing the degree of spatial correlation between neighboring elements on a 2-dimensional grid data from agricultural research, it calculates the following in its current form: 
$$I=\frac{N}{W}\frac{\sum_i \sum_j w_{ij}(z_i - \overline{z})(z_j - \overline{z})}{\sum_i (z_i - \overline{z})^2}$$
where $z_i$ is the value of random variable $z$ at each of the $N$ spatial locations, $w_{ij}$ is the weight between spatial locations $i$ and $j$, with $W=\sum_{i,j}w_{ij}$ and $\overline{z} = \sum_i z_i/N$. 
Moran's $I$ provides a global measure of whether the signed fluctuations away from the mean of quantity of interest $z$ at a pair of spatial locations correlates with the weight (frequently the inverse distance is used) between the locations. 
The metric found extensive use in fields that concern mapped data: econometrics\cite{anselin2001spatial}, ecology\cite{legendre1989spatial}, health sciences\cite{waller2004applied}, geology, and geography\cite{burrough2015principles}. 
Statistical distributions or their moments for Moran's $I$ under various conditions have been derived \cite{sen1976distribution,kelejian2001asymptotic,cliff-ord1973}. 
Moran's $I$, its local version, and the Moran scatterplot continue to find usage in geography and in fields dealing with mapped data.


\subsection{Geary's $C$}

Another early contender in the field is the Geary's $C$, originally named the contiguity ratio\cite{geary1954contiguity}. 
First demonstrated as a viable metric of spatial correlation on the example of demographic and agricultural data from counties of Ireland, it is defined:
$$C=\frac{N-1}{2W}\frac{\sum_i \sum_j w_{ij}(z_i - z_j)^2}{\sum_i (z_i - \overline{z})^2}$$ 
Moran's $I$ and Geary's $C$ have several features in common: both take the form of an outer product weighted by the spatial weights between the locations and both are normalized by the observed variance of $z$ and the sum of all spatial weights.
The distinction between them is the exact outer product operations carried out: Moran's $I$ multiplies the signed fluctuations away from the mean of $z$: $(z_i - \overline{z})(z_j - \overline{z})$, whereas Geary's $C$ takes the square of differences between values of $z$ at spatial locations $i$ and $j$: $(z_i - z_j)^2$. 
As such, Geary's $C$ takes on a large value for a variable that displays large variation among closely neighboring (large weight $w_{ij}$) spatial locations, whereas Moran's $I$ is large when the neighboring values fluctuate from the mean in the same direction. 
It is therefore natural that in the presence of multiple mapped variables, the scatter of $I$ vs. $C$ appear anti-correlated.

\subsection{Matheron's Variogram and $\gamma$}

Another metric is the variogram method of Matheron \cite{matheron1963principles} intended to quantify the typical variation of the spatial data points as a function of the distance separating them. 
Empirical variogram is often utilized in practice and is defined as follows: 
$$\hat{\gamma}(h \pm \delta) = \frac{1}{|N(h \pm \delta)|} \sum_{(i,j) \in N(h \pm \delta)} |z_i - z_j|^2$$
where $h$ is the distance between spatial locations with allowed tolerance $\delta$, $N(h \pm \delta)$ is set of all pairs of points $(i,j)$ such that distance between them lies in range $h \pm \delta$, and $z_i$ and $z_j$ are the values of the variable of interest at locations indexed $i$ and $j$, respectively.
Variogram analysis results in intuitive quantities: \textit{sill} and \textit{range} extracted from the curve of $\hat{\gamma}(h \pm \delta)$, where \textit{sill} indicates the eventual level of variability reached at asymptotic length scales, and \textit{range} denotes the length scale required to reach variability indistinguishable from the eventual \textit{sill}. 
Variogram is extensively used in geology as part of \textit{kriging} in mineral surveillance process\cite{davis1986statistics} and in atmospheric sciences\cite{nguyen2014satellite}.

\subsection{$\Gamma$ index and local $\Gamma$ index}

The global $\Gamma$ index, proposed in 1967, as a generalized method for identifying time-space clustering of cancer cases and other geographically labeled incidence data, tries to capture not only spatial, but also temporal information, albeit on binary variables indicating whether an incidence occurred or not \cite{mantel1967detection}. 
\lastcuts{It is therefore somewhat distinct from the other purely spatial metrics surveyed here. }
$\Gamma$ index considers two matrices: $a_{ij}$ and $b_{ij}$, one containing the measure spatial of similarity between incidences and the other containing the temporal similarity information. 
The statistic then is:
$$\Gamma = \sum_{i < j} a_{ij}b_{ij}$$
The local version of $\Gamma$ statistics avoids summation over index $i$, making it a quantity specific to the observation: $\Gamma_i = \sum_{j} a_{ij}b_{ij}$, and when summed equal to the global $\Gamma$, a property discussed in \ref{sec:lisa}. 

\subsection{Getis-Ord $G_i^*$}

\lastcuts{The 90's saw a rapid development in the field and what some termed the global-to-local transition\cite{getis2008history}.} 
Known as the $G$ and $G_i$ statistics, a class of metrics first formalized by Getis and Ord \cite{getis-ord1992} appeared as circumventing the shortcomings of the Moran's $I$ statistics. 
The local version of the statistic, $G_i$ is defined as follows: 
$$G_i(d) = \frac{\sum_{j \neq i} w_{ij}(d)z_j}{\sum_{j \neq i} z_j}$$
where $d$ is the length scale, the concentration of variable $z$ over which is being tested, and $w_{ij}$ is a binary weight indicating whether spatial locations $i$ and $j$ are within distance $d$ of each other. 
A minute modification of including the $i$th element in the summation over index $j$ turns the metric into $G_i^*(d)$. 
The global version of Getis-Ord statistics, measuring the overall level of concentration of variable $z$ in the neighborhood of linear scale $d$, is defined as follows:
$$G(d) = \frac{\sum_i \sum_j w_{ij}(d) z_i z_j}{\sum_i \sum_j z_i z_j}$$
The global $G(d)$ differs from Moran's and Geary's measures by taking the cross-product by multiplying the variables at locations $i$ and $j$ together: $z_i z_j$, instead of $(z_i - \overline{z})(z_j - \overline{z})$ in case of Moran's and $(z_i - z_j)^2$ in case of Geary's. 
For the $G_i^*$ statistic, a negative number reveals proximity of low values of the variable and a positive number - proximity of high values. 
The first empirical use case of $G$ statistics was to rule out significant spatial correlation on the county level data of Sudden Infant Death Syndrome for US state of North Carolina\lastcuts{ and to reveal concentration of low home values in San Diego} county beyond what Moran's $I$ would have indicated \cite{getis-ord1992}.

\subsection{Anselin's LISA and local Moran and Geary}
\label{sec:lisa}

Anselin proposed a generalized procedure for localizing the contribution of individual measurements on the global measure of spatial autocorrelation termed local indicators of spatial association (LISA). 
The method also serves to identify hot-spots or pockets of local variation in the mapped variable. 
LISA, broadly defined using two requirements: i) the statistic for a specific measurement should report whether similar values are clustered around it and ii) sum over all measurements should be proportional to a global statistic of spatial autocorrelation, generalizes the localized Moran's $I_i$ and Geary's $c_i$ statistics, also defined by Anselin \cite{anselin1995lisa}:
$$I_i = (z_i - \overline{z})\sum_j w_{ij}(z_j - \overline{z}) \text{ and } c_i = \sum_j w_{ij} (z_i - z_j)^2$$
Both local statistics are, in fact, proportional to their global counterparts with straight-forward proportionality constants, when summed up over all spatial locations. 
LISA's (specifically local Moran's $I_i$) first demonstrated usage was on dataset of international conflict among African nations, quantitatively identifying the hotbed of instability in Northeastern Africa. 
In the same category of techniques is the Moran's scatterplot, also outlined by Anselin\cite{anselin1996moran}, which disassociates low spatial autocorrelation into \textit{quadrants} of low values surrounded by high values and high values surrounded by low values, as well as high value of spatial autocorrelation into \textit{quadrants} of low values in among other low values and high values among other high values. 
See Getis\cite{getis2008history} for a
thorough history of spatial autocorrelation analysis.

\section{The \SkienaA{} Algorithm and Statistic}
\label{sec:method}

Our proposed method, which we term \SkienaA, produces a measure of spatial autocorrelation given a particular agglomeration order of $n$ locations $\{\hat{x}_i\}$ embedded in Euclidean space and values of random variable $\{z_i\}$ (with variance $\sigma^2$) paired with them. 
\SkienaA{} is agnostic to the exact clustering used, provided it is agglomerative and two clusters of spatial locations are merged at each step.

\SkienaA{} exploits the fact that the total sum of squared deviations ($SS(t)$) from the cluster mean of the variable $z_i$ increases monotonically as clusters are joined (proof in section \ref{sec:monotonicity}). 
This quantity is traced at a cost of constant time per merge event, starting when the first pair of observations are joined into a cluster and reaching $(n-1) \sigma^2$ when all observations are in a single cluster. 
We are interested in how quickly during the agglomeration process
this trace of sum of within-cluster squares takes off and reaches its eventual value of $(n-1) \sigma^2$. 


Formally, computation of \SkienaA{} starts with all coordinates as their own singleton clusters and keeps track of the geographic centroids of clusters ($\overline{\hat{x}}_{C_1}$ and $\overline{\hat{x}}_{C_2}$), their sizes ($|C_1|$ and $|C_2|$), means ($\overline{z}_{C_1}$ and $\overline{z}_{C_2}$), and the total sum of squares over all clusters: $SS(t) = \sum_{C_k \in C(t)} \sum_{i \in C_k} \left(z_i - \overline{z}_{C_k} \right)^2$ where $C(t)$ denotes the set of all clusters at time $t$ of the agglomeration order. 
During a merge event, clusters $C_1$ and $C_2$ are joined into a new cluster $C_{12}$ ($C_{12} \leftarrow C_1 \cup C_2$), with size $|C_{12}| \leftarrow |C_1| + |C_2|$, coordinate centroid
$$ \overline{\hat{x}}_{C_{12}} \leftarrow (|C_1|\overline{\hat{x}}_{C_1}+|C_2|\overline{\hat{x}}_{C_2})/|C_{12}| $$
and mean $\overline{z}_{C_{12}} \leftarrow (|C_1|\overline{z}_{C_1}+|C_2|\overline{z}_{C_2})/|C_{12}|$. 
The trace of sum of squares is updated as
$$ SS(t) \leftarrow SS(t-1) + |C_1|(\overline{z}_{C_{12}} - \overline{z}_{C_{1}})^2 + |C_2|(\overline{z}_{C_{12}} - \overline{z}_{C_{2}})^2 $$
It is then normalized by its final value, averaged over all agglomeration steps, and linearly transformed with $L(x)=2(1-x)-1$ to give the \SkienaA{} value:
$$ S_A = 2\left(1 - (\sum_{t \leq n - 1} SS(t))/((n-1) \cdot SS(n-1))\right) - 1 $$
with $n-1$ indicating the total number of merge events.

Just like conventional correlation coefficients, \SkienaA{} can range in the interval from -1 to 1. 
It will take 0 value when there is no spatial structure, larger value when similar values of $z_i$ are spatially nearby and negative values if neighboring values are anti-correlated.
\added{Intuitively, both nearby locations with very different values of feature $z_i$ and distant locations with similar values will decrease \SkienaA{}, while nearby locations with similar values and distant locations with differing values will contribute to the increase in \SkienaA{}.}
We note here that each update in the total sum of within-cluster squares due to a joining event is done in constant time, making calculation of \SkienaA{} for variable $z_i$ and any particular pre-specified agglomeration order an $O(n)$ algorithm. 
The required pre-computation of an agglomeration order can be performed in $O(n \log n)$ time, using single-linkage clustering in the plane. 
\swallow{
    We propose $S_A$ as a measure of the degree of spatial homogeneity present in mapped data, where each coordinate in 2-dimensional Euclidean space $(x_{i,1}, x_{i,2})$ is accompanied by a value $z_i$. $S_A$ is computed by way of establishing an agglomerative clustering scheme for the coordinates $(x_{i,1}, x_{i,2})$ defining the order in which clusters of $z_i$ values are formed and joined together. 
    By tracking the sum of squared deviations of $z_i$ values within clusters and combining over all clusters at a given time step during the agglomeration, we obtain a trace of the variability (Fig.~\ref{fig:trace}). 
    Averaging over all steps of the agglomeration and re-scaling by applying $S_A^{'} = 2(1-S_A)-1$ then provides a summary of the spatial similarity of $z_i$ values indicating how early or late in the agglomeration the correlation within the clusters disappeared. 
    Earlier decrease as in 2nd panel of Fig.~\ref{fig:trace} indicates that a variable like population density of US counties is not a spatially correlated variable, and that densely populated counties tend to neighbor sparse ones within small radius compared to the size of the US (Fig.~\ref{fig:mollweide}). 
    We note that $S_A$ is a quantity that is small when spatially nearby units have similar values (small intra-cluster variance at earlier stages) and large when nearby units are dissimilar in value. 
    Therefore, we introduce the rescaled/transformed version of Skiena-A, such that the quantity, intuitively, tracks the degree of autocorrelation of the variable, instead of its variance: $S_A^{'} = 2(1-S_A)-1$.
}

\subsection{Dependence on Agglomeration Order} 
\label{sec:linkage}

Multiple agglomerative clustering criteria are in common use, reflecting a trade-off between computational cost and robustness.
In this paper, we investigate four distinct criteria and their impact on observed spatial autocorrelations:

\begin{itemize}
    \item {\em Single linkage} --
    Here the distance between clusters $C_1$ and $C_2$ is defined by the closest pair of points spanning them:
    $$ d(C_1,C_2) = \min_{z_1 \in C_1, z_2 \in C_2} || z_1 - z_2 || $$
    This is akin to the criteria of Kruskal's algorithm for finding minimum spanning trees, and runs in $O(n \log n)$ time for the primary use case of points in the plane. 
    \added{The $O(n \log n)$ time is due to the disjoint set data structure with complexity bound of $O(\alpha(n))$ on merge/search operations. $\alpha$ is an extremely slowly increasing inverse Ackermann function and is a small constant for all practical purposes. }
    
    \item {\em Average linkage} --
    Here we compute distance between all pairs of cluster-spanning points, and average them for a more robust merging criteria than single-link:
    $$ d(C_1,C_2)  = \frac{1}{|C_1| |C_2|} \sum_{z_1 \in C_1} \sum_{z_2 \in C_2} || z_1 - z_2 || $$
    This will tend to avoid the skinny clusters of single-link, but at a greater computational cost.
    The straightforward implementation of average link clustering is $O(n^3)$, because each of the $n$ merges will potentially require touching $O(n^2)$ edges to recompute the nearest remaining cluster. 

    \item{\em Median linkage} -- 
    Here we maintain the centroid of each cluster, and merge the cluster-pair with the closest centroids. 
    The new merged cluster's centroid is given by the average of the centroids of the clusters being merged.
    This has two main advantages.
    First, it tends to produce clusters similar to average link, because outlier points in a cluster get overwhelmed as the cluster size (number of points) increases.
    Second, it is much faster to compare the centroids of the two clusters than test all $|C_1| |C_2|$ point-pairs in the simplest implementation.
    
    \item{\em Furthest linkage} --
    Here the cost of merging two clusters is the farthest pair of points between them:
    $$ d(C_1,C_2) = \max_{z_1 \in C_1, z_2 \in C_2} || z_1 - z_2 || $$
    This criteria works hardest to keep clusters round, by penalizing mergers with distant outlier elements. 
    Efficient implementations of furthest linkage clustering are known to run in $O(n^2)$ time. 
\end{itemize}

\begin{figure}[htbp]
    \begin{center}
        \includegraphics[width=\linewidth]{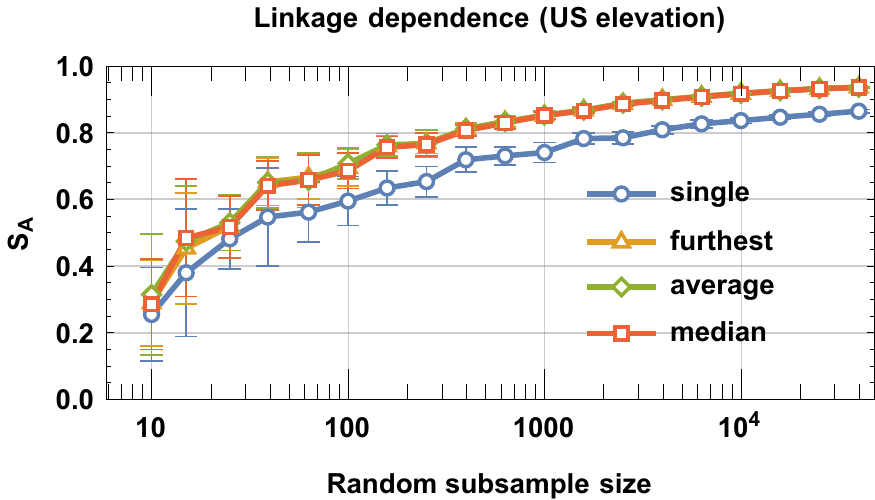}
    \end{center}
	\caption{\SkienaA{} calculated on subsample of the elevation data with different agglomeration methods. 
	All methods considered produce similar values of \SkienaA, except for single linkage.  }
	\label{fig:linkage}
\end{figure}

All linkage methods except for single linkage, produce similar results, while single linkage produces a slightly lower \SkienaA{} autocorrelation. 
This is natural as single linkage method merges only locally and suffers from what is known as the {\em chaining phenomenon} of returning spatially extent clusters with arbitrary shapes. 
The larger linear dimensions of the single linkage clusters reach the variability of the variable $z_i$ earlier driving the sum of squares up and the \SkienaA{} down (Fig.~\ref{fig:linkage}).

\subsection{Comparison with Moran's $I$ and Geary's $C$}

The comparison of median clustered \SkienaA{} with Geary's and Moran's can be seen in the scatter plot of the Fig.~\ref{fig:scatter} with each point representing a feature in the U.S. counties dataset. 
All 3 pairwise comparisons show large magnitude correlations $|r|>0.8$. 
In the bottom panel, single and median linkage methods are compared for \SkienaA. 
All panels use U.S. counties dataset with each point representing 2 different statistics computed for a particular column/feature of the dataset.

\section{Analysis of Statistical Properties}
\label{sec:analysis}

In this section, we prove three important properties of \SkienaA, namely monotonicity under merging, that it is a well-defined correlation measure with zero corresponding to no spatial correlation, and invariance under addition and multiplication by a constant.

\subsection{Monotonicity}
\label{sec:monotonicity}

For demonstration of the monotonicity of the total sum of within-cluster squared deviations from the mean of variable $z_i$, it suffices to show that an arbitrary cluster $C_1$ merging with another ($C_2$) would have non-decreasing squared deviation from the new cluster's mean $\overline{z}_{C_{12}}$ compared to the original mean $\overline{z}_{C_1}$.
Setting the mean shift equal to $\delta_z = \overline{z}_{C_{12}} - \overline{z}_{C_1}$, we compute the difference between the sum of square deviations from mean for $z_i$ values in cluster $C_1$ before and after the merge event as:
$$\sum_{i \in C_1}{(z_i - \overline{z}_{C_{12}})^2} - {(z_i - \overline{z}_{C_{1}})^2} = \overline{z}_{C_{12}}^2 - 2z_i\overline{z}_{C_{12}} - \overline{z}_{C_1}^2 + 2z_i \overline{z}_{C_{1}}$$
Substituting the mean shift $\delta_z$ and simplifying, we obtain:
$$\sum_{i \in C_1} 2 \overline{z}_{C_{1}} \delta_z + \delta_z^2 - 2 z_i \delta_z = \sum_{i \in C_1} \delta_z^2 = |C_1|\delta_z^2 \geq 0$$
where we have used the definition of mean to eliminate $z_i$ and $\overline{z}_{C_1}$. 
The change in sum of squared deviations for the clusters $C_1$ and $C_2$ being merged is, therefore, non-negative for all merge events, making the trace of $SS(t)$ a monotonic quantity. 
Its monotonicity, coupled with a suitable agglomeration order, which merges close-by coordinates earlier on, enables us to single out the area under its curve as a measure of spatial autocorrelation indicating how early/late in the agglomeration the variability increases from $0$ to $(n-1)\sigma^2$.

\subsection{Expected Value}

Intuitively, $S_A$ is \swallow{(linearly transformed with $L(x)=2(1-x)-1$)} mean of the (monotonically increasing) sum of squared deviations of values of $z_i$ from their cluster means while the observations are gradually merged into a single cluster made up of all coordinates $\hat{X}$. 
Under lack of spatial dependence, the sum of squared deviations will increase in even steps with no particular time structure and produce a mean over time equal to half its eventual value ($(n-1)\sigma^2/2$). 
After normalization and a linear transformation to flip the sign and adjust the range ($L(x)=2(1-x)-1$), we will obtain 0. 

For a formal proof, let us first consider $n$ real numbers $Z = \{z_1, ... z_n\}$ with mean $\overline{z}$ and Euclidean coordinates $\hat{X} = \{\hat{x}_1, ... \hat{x}_n\}$. 
Let $A(\hat{X}) = \{e_1, ... e_{n-1}\}$ a merge order that determines an agglomerative clustering on the symmetric weighted graph (with no self-edges) induced by a similarity metric on coordinates $\hat{X}$. 
Define the stages of this agglomeration at time $t$ as $A(\hat{X}, t) = \{e_1, ... e_t\}$ (with a shorthand $A(t)$) such that $A(\hat{X},n-1)=A(\hat{X})$. 
Let $C(t)$ denote the set of disjoint clusters present at time $t$ of agglomeration process such that $C(0)=\{\{1\},\{2\}, ... \{n\}\}$ and $C(n-1)=\{\{1,2, ... n\}\}$.

\begin{defn}
    $S_A$. 
    Define the $S_A$ statistic as: 
    $$ S_A(A(\hat{X}), Z) = 2\left(1 - \frac{\sum_{t = 1}^{n-1}SS(A(t), Z)}{(n-1) \sum_{i = 1}^{n} (z_i - \overline{z})^2} \right) - 1$$
    where $SS(A(t), Z) = \sum_{C_k \in C(t)} \sum_{i \in C_k} (z_i - \overline{z}_{C_k})^2$ (with a shorthand notation $SS(t)$) denotes the sum of within-cluster squared deviations at time $t$ of the agglomeration given by $A(t)$.
\end{defn}

\begin{thm}
    Let $Z = \{z_1, z_2, ... z_n\}$ be a set of normal i.i.d. random variables with mean 0 and variance $\sigma^2$ and $\hat{X}=\{\hat{x}_1, \hat{x}_2, ... \hat{x}_n\}$ their coordinates in Euclidean space. 
    Then the random variable $S_A(A(\hat{X}), Z)$ converges to zero in limit of large $n$:
    $$ \lim_{n \to \infty}\mathbb{E}[S_A(A(\hat{X}), Z)] = 0 $$
\end{thm}

\begin{figure}[htbp]
    \begin{center}
        \includegraphics[width=\linewidth]{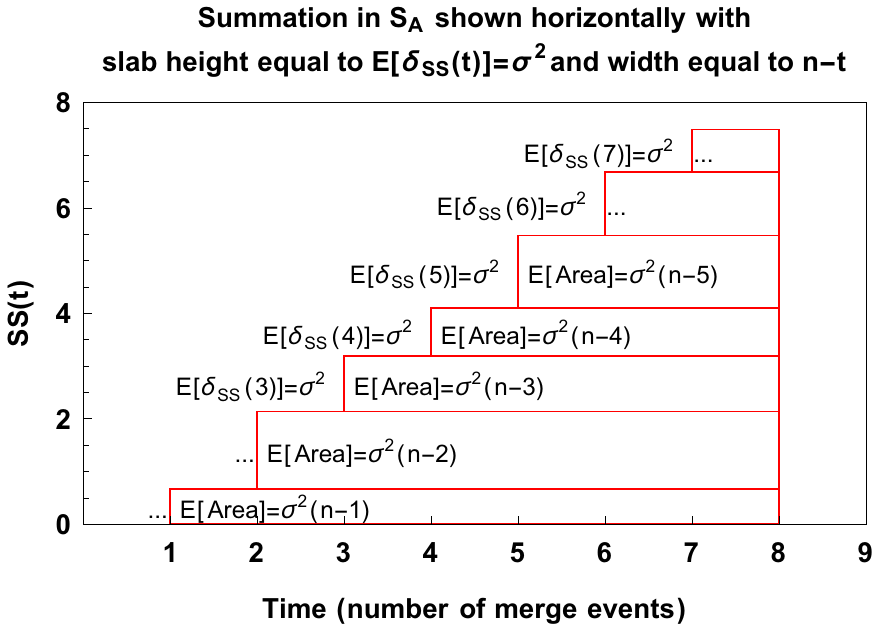}
    \end{center}
    \caption{Summation carried out in ``horizontal slabs", each with height in expectation equal to $\sigma^2$ and deterministic width of $n-t$.}
    \label{fig:proof}
\end{figure}

\begin{proof}
    We proceed by considering the contribution of each cluster joining event on the eventual metric $S_A$. 
    During a given merge event, clusters $C_1$ and $C_2$ with sizes $n_1$ and $n_2$ and means $\overline{z}_{C_1}$ and $\overline{z}_{C_2}$ join to make the cluster $C_{12}$ with size $n_{12} = n_1 + n_2$ and mean $\overline{z}_{C_{12}}$. 
    At the same time \lastcuts{, when $t \in \{0, 1, ... n-2\}$, between time steps $t$ and $t+1$}the running sum of within-cluster squares changes as follows (see Section \ref{sec:monotonicity}):
    \begin{align}
        \delta_{SS}(t+1)
        & = SS_{C_{12}}(t+1) - (SS_{C_1}(t) + SS_{C_2}(t)) \nonumber \\
        & = n_{1}(\overline{z}_{C_{12}} - \overline{z}_{C_{1}})^2 + n_{2}(\overline{z}_{C_{12}} - \overline{z}_{C_{2}})^2 \nonumber 
    \end{align}
    The expectation of change in sum of squared deviations due to merge event $\mathbb{E}[\delta_{SS}(t+1)]$ is then given by the difference in the expectations of sum of squares before and after the merge. 
    \begin{align}
        \mathbb{E}[\delta_{SS}(t+1)]
        & = \mathbb{E}[SS_{C_{12}}(t+1)] - (\mathbb{E}[SS_{C_1}(t)] + \mathbb{E}[SS_{C_2}(t)]) \nonumber \\
        & = (n_{12}-1)\sigma^2 - ((n_1-1)\sigma^2 + (n_2-1)\sigma^2) \nonumber \\
        & = (n_{12}-n_1-n_2+1)\sigma^2 = \sigma^2 \nonumber
    \end{align}
    Here we use the fact that for a given cluster $C$, $SS_{C}$ -- its sum of squared deviations from mean, is an estimate of the population variance biased by a factor of $n-1$. 
    The summation in definition of $S_A$ can then be carried out ``horizontally", by considering the jump in the global sum of squares times the number of time intervals for which this jump contributes to the metric as shown in Fig.~\ref{fig:proof}. 
    It then follows that: 
    \begin{align}
        \mathbb{E}[S_A(A(\hat{X}), Z)]
        & = 2 \left( 1 - \mathbb{E}\left[ \frac{\sum_{t = 1}^{n-1} SS(t)\lastcuts{\left(A(\hat{X}, t), Z\right)} }{ (n-1) SS(A(\hat{X}), Z) } \right] \right) - 1 \nonumber \\
        & = 2 \left( 1 - \frac{ \sum_{t = 1}^{n-1} \mathbb{E}\left[SS(t)\lastcuts{\left(A(\hat{X}, t), Z\right)} \right]}{ (n-1) SS(A(\hat{X}), Z) } \right) - 1 \nonumber \\
        & = 2 \left( 1 - \frac{\sum_{t = 1}^{n-1}(n-t)\mathbb{E}[ \delta_{SS}(t) ]}{(n-1) (n-1)\sigma^2} \right) - 1 \nonumber \\
        & = 2 \left( 1 - \frac{\left( (n-1)n-(n-1)n/2\right)\sigma^2)}{ (n-1)(n-1)\sigma^2}\right) - 1 \nonumber \\
        & = - \frac{1}{n-1} \nonumber
    \end{align}
    Here we use the fact that the distribution of overall sum of squares in the denominator is related to the sampling distribution of sample variance: 
    $$\frac{SS(A(\hat{X}), Z)}{\sigma^2} = \frac{\sum_{i = 1}^{n} (z_i - \overline{z})^2}{\sigma^2} \sim \chi^2(n-1)$$
    making $SS(A(\hat{X}), Z)$ a self-averaging quantity with mean $(n-1)\sigma^2$ and variance $2(n-1)\sigma^4$, and hence vanishing relative variance in the limit of large $n$: 
    $$\lim_{n \to \infty} \frac{Var[SS(A(\hat{X}), Z)]}{\mathbb{E}[SS(A(\hat{X}), Z)]^2} = \lim_{n \to \infty} \frac{2(n-1)\sigma^4}{(n-1)^2\sigma^4} = 0$$
    This lets us treat $SS(A(\hat{X}), Z)$ in denominator as a constant factor and taking the limit of large $n$ of $\mathbb{E}[S_A(A(\hat{X}), Z)]$, we obtain:
    $$ \lim_{n \to \infty}\mathbb{E}[S_A(A(\hat{X}), Z)] = \lim_{n \to \infty} \left( - \frac{1}{n-1} \right) = 0 $$ 
    as desired. 
\end{proof}

\subsection{Invariance}

\begin{figure}[htbp]
    \begin{center}
        \includegraphics[width=\linewidth]{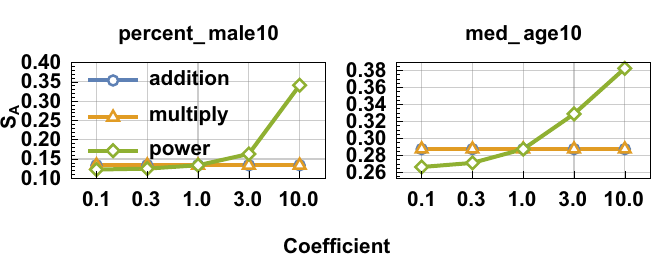}
    	\caption{Invariant properties of \SkienaA: $S_A(a+X) = S_A(X)$ and $S_A(a \cdot X) = S_A(X)$, but $S_A(X^a) \neq S_A(X)$.}
    	\label{fig:math-properties}
    \end{center}
\end{figure}

The \SkienaA{} statistic has the nice property of invariance under addition and multiplication by a constant.
Letting $Z$ a spatial variable with $S_A(Z) = s$ and considering $S_A(Z+c)$ with $c \in \mathbb{R}$, we note that the sum of squared deviations is unaffected by addition of a constant, making our statistic invariant to addition of a constant $c$.
\begin{align}
    SS(T(t, \hat{X}), Z+c) 
    & = \sum_{\substack{C_k \in C(t) \\ e_i \in C_k}} \left(z_i + c - \frac{\sum_{e_j \in C_k} z_j + c}{|C_k|} \right)^2 \nonumber \\
    & = \sum_{\substack{C_k \in C(t) \\ e_i \in C_k}} \left(z_i - \frac{\sum_{e_j \in C_k} z_j}{|C_k|} \right)^2 \nonumber \\
    & = SS(T(t, \hat{X}), Z) \nonumber
\end{align}
Considering multiplication of variable $Z$ by an arbitrary constant $c \in \mathbb{R}$, we note that a factor of $c^2$ appears both in denominator and numerator due to the squared deviation from the mean being considered, canceling each other and returning the same value as the original variable $S_A(c \cdot Z) = S_A(Z)$. 
Fig.~\ref{fig:math-properties} illustrates this property for two demographic variables, and shows that it does not hold for exponentiation.

\section{Experimental Evaluation}
\label{sec:evaluation}

Here we present the results of simulations which demonstrate
(1) the running time of \SkienaA{} is indeed an order of magnitude faster to compute than competing statistics,
(2) \SkienaA{} identifies substantially weaker spatial correlations in synthetic data than Moran's and Geary's statistics,
(3) \SkienaA{} appears to be influenced less by non-uniform sampling than competing statistics, and finally
(4) \SkienaA{} appropriately reports increased autocorrelation with greater sampling density while still converging to a limit below the perfect autocorrelation of 1.

\subsection{Running Time}

\begin{figure}[htbp]
    \begin{center}
	    \includegraphics[width=\linewidth]{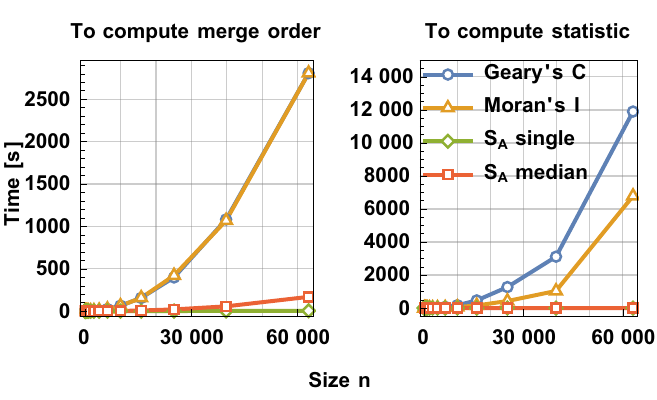}
    \end{center}
	\caption{Experiments concerning running time. 
	Single-link and median-link agglomeration orders cost less to compute than single runs of Moran $I$ and Geary's $C$ metrics. 
	\SkienaA{} outperforms $I$ and $C$ drastically given the merge order on a dataset of size $\approx 63000$.}
	\label{fig:timing}
\end{figure}

\SkienaA{} substantially outperforms both Moran's and Geary's metrics in computation time, both in establishing the agglomerative merging order to use and to compute the statistics.
In our experiments, computing a single median-linkage agglomeration order costs approximately 10\% of a {\em single} Moran or Geary computation on the same points, as shown in Fig.~\ref{fig:timing} (left).
By reusing this agglomeration order we can save a linear factor of running time on subsequent autocorrelation analyses.
Fig.~\ref{fig:timing} (right) shows that for a typical dataset of $n \approx 63,000$ points, our $S_A$ autocorrelation measure can be computed in 1 second, versus 2 hours or more for Moran's $I$ and Geary's $C$.

Timing experiments were done as follows: starting from coordinates, agglomeration order was computed using Kruskal's routine with disjoint set structure (for \SkienaA{} single), scipy's linkage tool (for \SkienaA{} median) and numpy's linear algebra toolbox with vectorization (for weight matrix of Moran's $I$ and Geary's $C$) and metrics were computed using our streaming tool (\SkienaA{}) and pysal library for python (Moran's $I$ and Geary's $C$). 
All tools were written in python 3.7.

\subsection{Reusing agglomeration order: fMRI time series analysis}

Much of the efficiency gains \SkienaA{} accrue from its ability to reuse a once-computed agglomeration order for new data points arriving from the same spatial coordinates. 
We demonstrate this with an application to functional neuroimaging data (fMRI), which gives a time series readout for each spatial location in the brain. 
In order to study the dynamics of brain networks, neuroscience is concerned with extracting summary statistics from the brain images of potentially $>10^6$ voxels (3D pixels) at the resolution of sampling period. 
The statistics are then used in downstream prediction and classification tasks of clinical significance. 
In this experiment, we used a publicly available fMRI neuroimaging dataset with 36 fMRI scans (12 human subjects $\times$ 3 experimental conditions) with each scan consisting of $2 \times 2,320 = 4,640$ repeated measurements of the entire brain at $0.8s$ sampling period \cite{pnas-diet}. 
We focused on the grey matter data, which consists of readings from $n=133,000 \pm 13,000$ (mean $\pm$ std) voxels at each time point. 
To compute \SkienaA{}, we constructed a single agglomeration order for each scan, using k-d tree structure by treating the grey matter voxels of brain as points in space to be partitioned into singletons.
We cycled through the three axes of brain recursively, splitting each partition between its median pair of planes perpendicular to the axis until all partitions reached size of 1. 
The splitting events then define an agglomeration order in reverse.
The time complexity of partitioning space using k-d tree structure is $O(n)$ in case of unbalanced tree, and $O(n\log n)$ for a balanced tree with median finding subroutine. 
%
Due to the highly irregular shape of the grey matter, we resorted to finding the medians for balanced partitions, with the average time to establish the agglomeration order of two minutes, but it can be reused for each of the $m$ time points of a given scan.
This reduces the run time from $O(mn^2)$ for Moran's $I$ and Geary's $C$ to $O(n \log n + mn)$. 
In our case, with $m=4640$ time points and $n\approx 133000$ coordinates, \SkienaA{} took $3500 \pm 300$ seconds, or $0.75 \pm 0.07$ seconds per feature (time step). 
On the other hand, we were not able to compute Moran's $I$ and Geary's $C$ for $133,000$ coordinates on an average workstation hardware using the standard implementation (pysal), due to space limitations. 
We give a linear-time algorithm to calculate $S_A$ for a variable with an input agglomeration order (available at \putlink{}).

\begin{figure}[htbp]
    \begin{center}
	    \includegraphics[width=\linewidth]{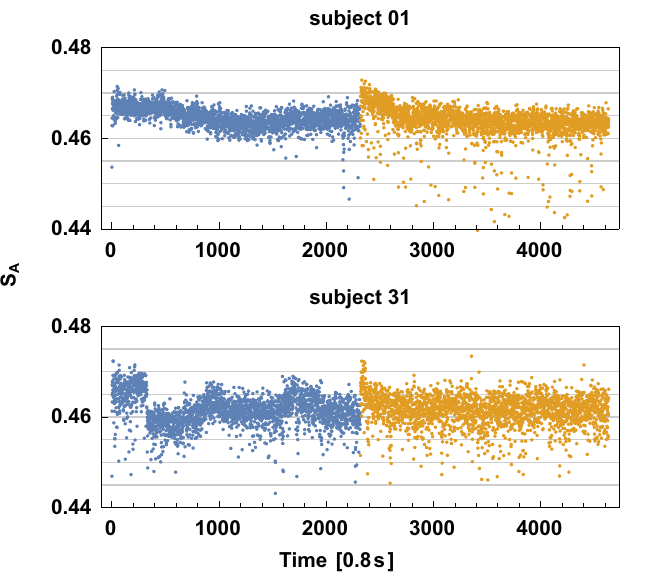}
    \end{center}
	\caption{Spatial autocorrelation (measured by $S_A$) time series for fMRI data, showing visibly different degrees of coherence on two different human subjects.  
	We estimate that this computation would have taken roughly 36,000 times as long using either the Moran's I or Geary C statistic.  
	The two colors indicate the two halves of the scanning session, with short break in the middle. }
	\label{fig:brain-peak-timeseries}
\end{figure}

Extrapolation from computations of Moran's $I$ and Geary's $C$ on smaller samples indicate that if memory requirements were lifted, it would take more than $7.5$ and $13.5$ hours respectively for each time step of the time series data, or roughly $36,000$ times longer than \SkienaA{}. 
Fig~\ref{fig:brain-peak-timeseries} shows representative autocorrelation time series from brain fMRI data.
This shows that \SkienaA{} not only improves computation for each data feature, but also processes each additional feature in linear time by reusing the agglomeration order once it is computed.
\SkienaA{}'s complexity for each time step is comparable to the sampling period of the fMRI data. 
This permits future applications in closed-loop systems that process data and provide feedback stimuli or electromagnetic stimulation to the brain in real-time for improved clinical intervention.

\subsection{Sensitivity to True Autocorrelation: Synthetic Data}

Ground truth on the degree of spatial autocorrelation can only be obtained from simulation results, where we explicitly generate data with specified amount of spatial autocorrelation and see how much bias must be added for statistics to identify the phenomenon.
For this purpose, we carry out a disk-averaging experiment, whereby a normally distributed independently sampled random variable $z_i$ is assigned to uniformly distributed coordinates and undergoes an averaging procedure. 
The averaging takes all values of $z_j$ for locations within disk of radius $r$ around coordinate $\hat{x}_i$, and reassigns the average of the within disk values to it: $z_i \leftarrow mean(\{ z_j \,|\, d(\hat{x}_i - \hat{x}_j) < r \})$.
The \SkienaA{} statistic of the disk-averaged $z_i$ values were computed and compared to Moran's $I$ and Geary's $C$.
Random sampling, disk-averaging and statistic computation were each repeated 100 times. 

\begin{figure}[htbp]
    \begin{center}
	    \includegraphics[width=\linewidth]{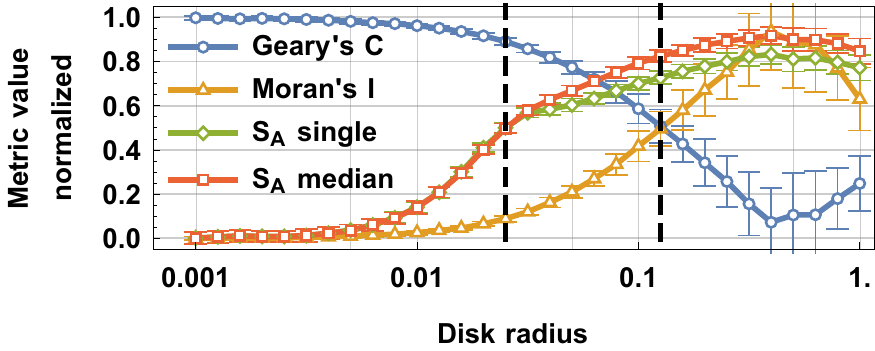}
    \end{center}
	\caption{$S_A$ is more sensitive to true autocorrelation than Moran's $I$ and Geary's $C$, on a ``disk-averaging'' generative model as a function of disk radius. 
	Moran's $I$, Geary's $C$ values are rescaled to match the range of $S_A$. 
	$S_A$ detects the autocorrelation $>0.5$ order of magnitude earlier than Moran's $I$ and Geary's $C$. 
	Note the entire range of $[0, 0.9]$ is covered with \SkienaA{} within 2 orders of magnitude of the disk radius. 
	Vertical dashed lines indicate disk radii where metrics reach half of their ranges.}
	\label{fig:disk-mean}
\end{figure}

Fig.~\ref{fig:disk-mean} summarizes the results of these experiments for $1000$ points. 
\SkienaA{} (both single and median-linkage) demonstrates far greater sensitivity, identifying significant and rapidly increasing amounts of spatial autocorrelation for disk radii half an order of magnitude smaller than that of Geary's $C$ and Moran's $I$. 
Although both Moran and Geary statistics support problem-specific weight matrices to tune their sensitivity, the interesting autocorrelation distance scales are a priori unknown and difficult to determine, so methods without tunable parameters are preferred.

\swallow{
    \begin{algorithm}
        \SetAlgoLined
        \SetKwData{Radius}{r}\SetKwData{CountM}{m}\SetKwData{Zdata}{$z$}\SetKwData{Zr}{$z_r$}\SetKwData{Coordinate}{$\hat{x}$}\SetKwData{Disk}{disk}
        \SetKwFunction{Mean}{mean}\SetKwFunction{Distance}{d}\SetKwFunction{Uniform}{Uniform}
        \SetKwInOut{Input}{input}\SetKwInOut{Output}{output}
        \Input{radius of disk \Radius}
        \Input{number of samples \CountM}
        \Output{Spatial data \Zr with spatial correlation given by radius \Radius}
        \For{$i\leftarrow 1$ \KwTo \CountM}{
            $\Zdata[i] \leftarrow$ random sample from $\mathcal{N}(0,1)$\;
            $\Coordinate[i] \leftarrow$ random sample from \Uniform{$[0,1]^2$}\;}
        \For{$i\leftarrow 1$ \KwTo \CountM}{
            $\Zr[i] \leftarrow$ \Mean{$\{\Zdata[j] \,|\, \Distance{\Coordinate [i] - \Coordinate [j]} < \Radius\}$}\;}
        \caption{Disk-averaging algorithm for inducing spatial autocorrelation on synthetic data}
        \label{algo:disk-average}
    \end{algorithm}
}

\subsection{Sensitivity to Sample Size and Coordinate Subsampling: U.S. Elevation Data}

\begin{figure}[htbp]
    \begin{center}
        \includegraphics[width=\linewidth]{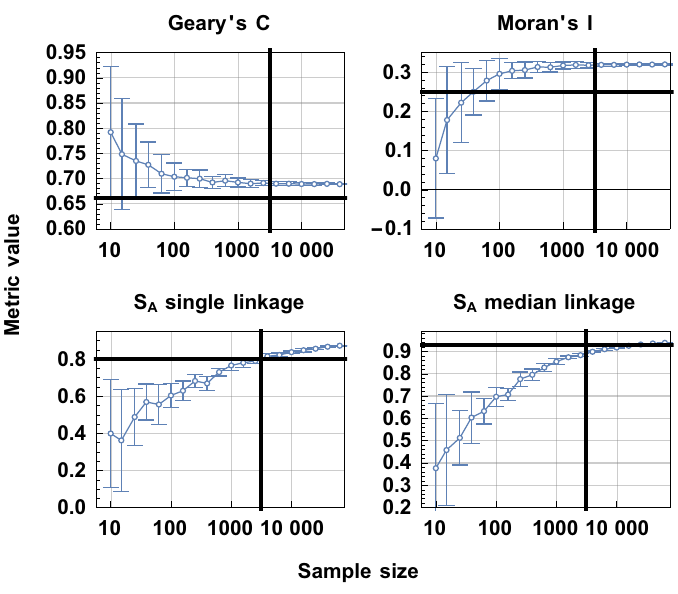}
    \end{center}
	\caption{
	$S_A$ reveals autocorrelation independent of the exact coordinates sampled. 
	The random subsampling experiment on $1km^2$ scale US elevation data carried out up to subsample size 40000. 
	Vertical and horizontal lines indicate the number of counties in the U.S. counties dataset and the value of metric computed from them, respectively.}
	\label{fig:random-subsample}
\end{figure}

Spatial autocorrelation depends on the exact sampling of the coordinates as well as the spatial distance/weight matrix. 
We note that for historical and demographic reasons, U.S. counties are not of equal size and shape, but generally smaller and more irregular in the east rather than the west.
A spatial autocorrelation statistic should ideally report similar values on the same underlying geographic variable regardless of the details of the sampling method.

To interrogate whether \SkienaA{} computed on subsamples of real data differs from Moran and Geary's statistics in its dependence on the exact subsample of coordinates, we use the following procedure. 
$n$ random data points are drawn from the U.S. elevation data (itself sampled at $1km^2$) \cite{2007-NA-1km-grid}, and \SkienaA, Moran's $I$ and Geary's $C$ are computed from their coordinates $\hat{x}_i$ and elevation values $z_i$. 
Performing the experiment at sample sizes up to 40,000 points (limited by the $O(n^2)$ running time of Moran and Geary's), we compute autocorrelation metrics, and compare them to the values obtained from the elevation column of the U.S. counties dataset at sample size of 3142. 
Results shown in Fig.~\ref{fig:random-subsample}.


\swallow{
    \begin{algorithm}[h]
        \SetAlgoLined
        \SetKwData{Nsamples}{m}
        \SetKwData{choice}{choice}
        \SetKwData{Xelev}{$\hat{x}_{elev}$}
        \SetKwData{Zelev}{$z_{elev}$}
        \SetKwData{Xsample}{$\hat{x}$}
        \SetKwData{Zsample}{$z$}
        \SetKwData{SAvalue}{$S_A$}
        \SetKwFunction{SA}{\SkienaA}
        \SetKwFunction{RandChoice}{RandChoice}
        \SetKwInOut{Input}{input}\SetKwInOut{Output}{output}
        \Input{\Zelev (U.S. elevation data at $1km^2$ resolution)}
        \Input{\Nsamples (number of random samples to take from U.S. elevation data)}
        \Output{$S_A$ (\SkienaA{} value from U.S. elevation data \Zelev at \Nsamples samples)}
        \For{$i \leftarrow 1$ \KwTo \Nsamples}{
            $\choice \leftarrow$ \RandChoice{$\{1, ... | \Xelev |\}$}\;
            $\Xsample[i] \leftarrow \Xelev[\choice]$\;
            $\Zsample[i] \leftarrow \Zelev[\choice]$}
        \SAvalue $\leftarrow$ \SA{\Xsample, \Zsample}
        \caption{Routine for subsampling elevation on the continental U.S. data.}
        \label{algo:elev-subsample}
    \end{algorithm}
}

Both Moran's $I$ and Geary's $C$ report different values when the coordinates are sampled uniformly, compared to the irregular sample of coordinates given by U.S. counties' locations.
On the other hand, both single- and median-linkage \SkienaA{} report similar values with equal number of uniformly sampled coordinates as it did with coordinates of U.S. counties, showing robustness to changes in the exact subsampling of coordinates. 


\subsection{Convergence Evaluation and Analytical Fit}

To test convergence of \SkienaA, Moran's, and Geary's metrics we perform the following sampling procedure on grids of random values of varying sizes. 
For a rectangular grid of finite size e.g. $k \text{-by-} k$, we assign a uniformly random $z_{ij}$ value to each of the $k^2$ grid cells, then randomly sample $n$ real valued coordinates from the support given by $[0,k]^2$, and take their corresponding cell's $z_{ij}$ values to compute \SkienaA{}. 
This procedure locks a particular correlation length into the data by choosing the number of grid cells, and forces the metrics to capture it as number of sample coordinates increases.  
We expect $1/k^2$th of all samples to fall in each grid cell, thus taking on the same $z$ value, and raising the autocorrelation as the number of samples increases to a natural limit, because there will also be nearby pairs of points that sit across a grid boundary and take different $z$ values.
Thus a meaningful metric should converge to a large value (but less than the maximum possible 1) that decreases for shorter autocorrelation lengths induced by larger number of grid cells.

\begin{figure}[htbp]
    \begin{center}
        \includegraphics[width=\linewidth]{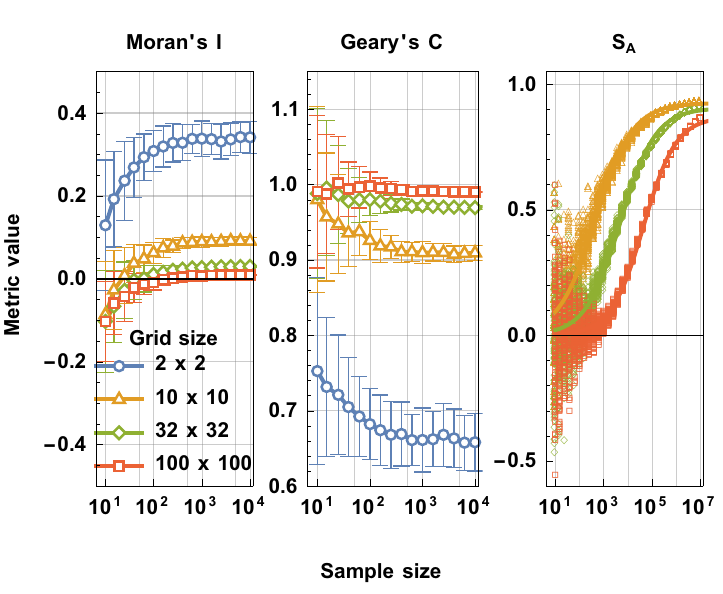}
    \end{center}
	\caption{Asymptotic behavior of spatial correlation metrics. 
	Random coordinates are sampled at increasing sample size from square grid of independent random values from [0, 1] interval. 
	\emph{Left}: Moran's $I$, \emph{center}: Geary's $C$, and \emph{right}: \SkienaA, solid lines represent best fit of log-transformed sigmoid curves for data drawn from grids of size 10 x 10. 
	$\frac{0.925}{1+\exp(-0.567(\log N - 6.137))}$, 32 x 32: $\frac{0.905}{1+\exp(-0.6(\log N - 8.522))}$, 100 x 100: $\frac{0.87}{1+\exp(-0.644(\log N - 10.722))}$.
	Note the asymptote of single-linkage \SkienaA{} converging to values $< 1$: $S^{10}_{max}=0.925$, $S^{32}_{max}=0.905$, and $S^{100}_{max}=0.87$.}
	\label{fig:grid-sample}
\end{figure}

Fig.~\ref{fig:grid-sample} (left) reports that Moran's $I$ converges to values increasingly closer to $0$ as the grid size increases, indicating it captures the de-correlated structure of large number of random grid cell entries $z_{ij}$. 
Geary's $C$ does similarly, reporting values increasingly closer to $1$. 
But \SkienaA{} clearly \emph{sees} the coarser, more correlated structure of smaller grids with fewer samples, reporting \emph{earlier} increase for $10 \text{-by-} 10$ grid than for $100 \text{-by-} 100$ (Fig.~\ref{fig:grid-sample}, right panel). 

\swallow{
    \begin{algorithm}
        \SetAlgoLined
        \SetKwData{Ngrid}{n}
        \SetKwData{Nsamples}{m} 
        \SetKwData{Xgrid}{$\hat{x}_{grid}$}
        \SetKwData{Xsample}{$\hat{x}$}
        \SetKwData{Zgrid}{$z_{grid}$}
        \SetKwData{Zsample}{$z$}
        \SetKwData{SAvalue}{$S_A$}
        \SetKwData{Disk}{disk}
        \SetKwFunction{SA}{\SkienaA}
        \SetKwFunction{Uniform}{Uniform}
        \SetKwFunction{Ceiling}{Ceiling}
        \SetKwFunction{Mean}{mean}
        \SetKwFunction{Distance}{d}
        \SetKwInOut{Input}{input}\SetKwInOut{Output}{output}
        \Input{\Ngrid (number of grid cell along each axis)}
        \Input{\Nsamples (number of samples)}
        \Output{$S_A$ (\SkienaA{} value for grid size $\Ngrid^2$ at \Nsamples samples)}
        \For{$i \leftarrow 1$ \KwTo \Ngrid}{
            \For{$j \leftarrow 1$ \KwTo \Ngrid}{
                \Zgrid$[i,j] \leftarrow$ sample from \Uniform{$[0,1]^2$}\;}}
        \For{$i\leftarrow 1$ \KwTo \Nsamples}{
            $\Xsample[i] \leftarrow$ sample from \Uniform{$(0,\Ngrid]^2$}\;
            $\Zsample[i] \leftarrow$ \Zgrid [\Ceiling{$\Xsample[i,1]$}, \Ceiling{$\Xsample[i,2]$}]\;}
        \SAvalue $\leftarrow$ \SA{\Xsample, \Zsample}
        \caption{Routine for randomly sampling from a grid of values.}
        \label{algo:grid-sample}
    \end{algorithm}
}

\begin{figure}[htbp]
    \begin{center}
        \includegraphics[width=\linewidth]{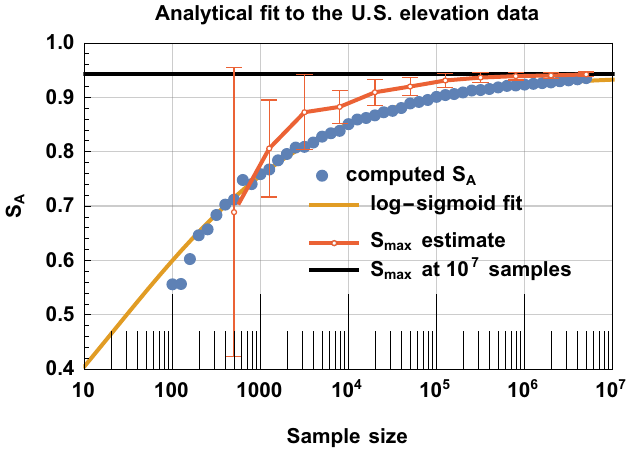}
    \end{center}
	\caption{Estimation of $S_{max}$. 
	Yellow curve: best fit to \SkienaA{} as a function of sample size $S_A(N)=\frac{0.942}{1+\exp(-0.361(\log N - 3.035))}$ using data points in blue. 
	Red curve: the confidence interval of parameter estimation for the asymptotic value $S_{max}$ using the \SkienaA{} computed only up to the sample size on the $x$-axis. U.S. elevation data.}
	\label{fig:skienaA-elevation-fit-quality}
\end{figure}

In order to estimate the asymptotic value of the $S_A$ metric, we fit the following log-sigmoidal functional form to the observed values of $S_A$ as a function of samples taken: $S_A(n) = S_{max}/(1+e^{-a(\log n - b)})$.
The parameter $S_{max}$ has a natural interpretation of the asymptotic value of $S_A$ at unlimited number of samples, turning the task of finding the asymptote into a parameter estimation for $S_{max}$.
See Fig.~\ref{fig:skienaA-elevation-fit-quality}.
We report that with sample size $>10^5$, the confidence interval for estimated $S_{max}$ includes the eventually best estimate (black line) computed using $10^7$ samples. 
None of the estimates of $S_{max}$ includes the value of 1. 

\section{Conclusion}

The Skiena's A (\SkienaA{}) algorithm and statistic we propose provides an efficient, improved sensitivity procedure for computing the spatial autocorrelation, running in linear time after computing the agglomeration order (implementation available at \putlink{}).
Separating the computation into two steps: i) obtaining the agglomeration order and ii) computing of the statistic, provides additional improvements by reusing the agglomeration order for new data that arrive from the same coordinates. 
\SkienaA{} achieves run time of $O(n \log n + mn)$ for $m$ separate features, improving upon the standard $O(mn^2)$. 
As demonstrated in the fMRI example, it can be thousands of times faster in natural time series applications of spatial autocorrelation than previous methods.
Even for single-shot applications in the plane where we can compute single-linkage agglomeration in $O(n \log n)$ run time, we beat previous $O(n^2)$ algorithms.
\added{We envision that a natural domain of application of our method will be in tracking the time-dependence of spatial similarity of features in spatially-tagged data. }
We have also shown that \SkienaA{} has the convenience of converging to 0 for random data, invariance under linear transforms uniformly applied to data, making it an attractive addition to standard toolbox for analysis of spatial data irrespective of the domain.

\bibliographystyle{IEEEtran}
\bibliography{ms}

\end{document}